# Interaction of fibrinogen–magnetic nanoparticle bioconjugates with integrin reconstituted into artificial membranes†

Ulrike Martens, ‡[a,b] Una Janke,‡[a,b] Sophie Möller,[a,b] Delphine Talbot,[c] Ali Abou-Hassan *[c] and Mihaela Delcea *[a,b]

Magnetic nanoparticles have a broad spectrum of biomedical applications including cell separation, diagnostics and therapy. One key issue is little explored: how do the engineered nanoparticles interact with blood components after injection? The formation of bioconjugates in the bloodstream and subsequent reactions are potentially toxic due to the ability to induce an immune response. The understanding of the underlying processes is of major relevance to design not only efficient, but also safe nanoparticles for *e.g.* targeted drug delivery applications. In this study, we report on maghemite nanoparticles functionalized with citrate-, dextran- and polyethylene glycol coatings and their interaction with the clotting protein fibrinogen. Further, we investigate using biophysical tools (*e.g.* dynamic light scattering, circular dichroism spectroscopy and quartz crystal microbalance) the interaction of the magnetic nanoparticles–fibrinogen bioconjugates with artificial cell membranes as a model system for blood platelets. We found that fibrinogen corona formation provides colloidal stability to maghemite nanoparticles. In addition, bioconjugates of fibrinogen with dextran- and citrate-coated NPs interact with integrin-containing lipid bilayer, especially upon treatment with divalent ions, whereas PEG-coating reveals minor interaction. Our study at the interface of protein-conjugated nanoparticles and artificial cell membranes is essential for engineering safe nanoparticles for drug delivery applications.

## Introduction

Among the properties of magnetic nanoparticles (MNPs), which paved their way in the field of biomedical applications, is their physical ability to respond to a magnetic field.[1–5] In addition to this property, magnetic NPs can be specifically engineered with a desired chemical composition, shape, roughness, size, surface charge and coatings to meet the needed requirements for a specific application. The majority of applications of these NPs including *e.g.* imaging and targeted drug delivery involves blood contact. Once the NPs enter the body, they can trigger different reactions due to their interplay with blood components.[6] In blood, proteins attach immediately to the nanoparticle surface forming a so-called "protein corona", which defines the biological identity of the colloidal suspension. Such processes in the blood stream can lead to alterations of protein structures which may cause diseases through the ability to induce an immune response.[6,7] Additionally, the study of potential interaction of nanoparticles with cell membrane proteins is crucial for the design and development of highly effective and non-toxic NPs for biomedical applications.

When NPs reach the target cells, they may either show no interaction with the cell membrane or can adhere to the membrane being further wrapped and subsequently internalized.[8,9] Naked NPs with a high surface energy show strong adsorption to membranes as a result of unspecific interactions. Consequently, surface modification and corona formation lower their surface energy restricting the unspecific binding.[10,11] Lesniak *et al.* described the NP-membrane interaction as a two-step process starting with NP adhesion to the cell membrane and interaction with proteins and lipids followed by an energy-dependent internalization mechanism of the cell.[11] However, the scenario describing the biological system of nanoparticles and membranes interaction is much more complex. The NP corona not solely consists of abundant proteins; it comprises also complement proteins that activate the innate immune system. Therefore, opsonization and subsequent phagocytosis by immune cells are also present.[12] In addition, internalization of NPs can occur indirectly through diffusion, where NPs can penetrate cells by passive translocation or endocytosis pathways. A detailed description of interaction mechanisms can be found elsewhere.[13,14]

[a]Institute of Biochemistry, University of Greifswald, 17489 Greifswald, Germany. E-mail: delceam@uni-greifswald.de
[b]ZIK HIKE – Center for Innovation Competence "Humoral Immune Reactions in Cardiovascular Diseases", University of Greifswald, 17489 Greifswald, Germany
[c]Sorbonne Université, CNRS UMR 8234, Physico-chimie des Électrolytes et Nanosystèmes InterfaciauX, F-75005 Paris, France. E-mail: ali.abou_hassan@sorbonne-universite.fr

† Electronic supplementary information (ESI) available. See DOI: 10.1039/d0nr04181e

‡ Both authors contributed equally

Fibrinogen, is a 340 kDa plasma protein with three disulfide-bonded chains (α-, β- and γ-chain) that are linked together by a dimeric disulfide knot (DSK) at the *N*-terminus. It is a highly abundant blood protein with a concentration in human plasma of about 2.0–4.5 mg $mL^{-1}$.[15–17] Fibrinogen forms polymeric fibrin responding to injuries of the vascular system, which is important for the clotting process and platelet aggregation. Additionally, fibrinogen binds another major player important for haemostasis and cell adhesion, which is the heterodimeric platelet receptor integrin αIIbβ3. This 235 kDa bidirectional receptor undergoes "outside-in" and "inside-out" signalling leading to high- or low-affinity conformations as demonstrated by several techniques, *e.g.* cryo-electron microscopy (EM) and negatively stained EM.[18] Agonist-activation of platelets results in the opening of the integrin head domain (inside-out signalling), which enables fibrinogen binding *via* its RGD (Arginine–Glycine–Aspartate peptide) motifs in the Aα chain or by the KQAGDV sequence located in the γ-chain.[19] As a consequence, intracellular signalling pathways are activated and thus, platelet aggregation (outside-in signalling) is induced.[20] Moreover, fibrinogen binding involves the metal-ion-dependent adhesion site (MIDAS), the adjacent MIDAS (ADMIDAS) and the synergistic metal ion binding site (SyMBS) of the integrin, that depend on divalent cations. $Mg^{2+}$, $Ca^{2+}$ and $Mn^{2+}$ have been widely shown to be crucial for integrin stability, fibrinogen binding and its activation state. Especially binding of $Mn^{2+}$ has been discussed to promote a structural change towards the active integrin.[21,22]

Reactions between the proteins immobilized on the surface of the NPs and the membrane of the platelets may cause structural changes of the involved proteins or alterations in the clotting process which are not intended, and thus, are of potential risk.[23,24] Consequently, it appears clearly that the elaboration of a mimetic model where the interaction between NPs and the proteins of the membrane can be studied using biophysical tools will advance our understanding of the subject.

One parameter affecting the interaction of NPs with *e.g.* blood components is their surface functionalization.[25] Here, we report on the interaction of citrate-, dextran- and polyethylene glycol (PEG) surface-modified maghemite (γ-$Fe_2O_3$) NPs with the abundant blood protein fibrinogen (Fb) and the characterization of the interface of Fb-conjugated maghemite NPs with an artificial cell membrane containing αIIbβ3 integrin as a model system for blood platelets. The surface modifications of NPs were characterized using Fourier transform infrared spectroscopy (FTIR) and dynamic light scattering (DLS). The successful formation of the Fb corona on the NPs surface was demonstrated by DLS measurements as well as SDS-PAGE. Further, the interaction of NP–Fb bioconjugates with αIIbβ3-containing lipid membranes was studied using quartz crystal microbalance with dissipation monitoring (QCM-D) technique. The transmembrane protein was activated by the addition of divalent ions. This allows a comparison of the interactions of NPs with integrin in active- or inactive state. We found that fibrinogen-functionalized NP-integrin interaction plays an important role in mimicking platelet activation by αIIbβ3 conformational change *via* divalent ions.

## Experimental

### Fibrinogen and others

Iron(III) nitrate nonahydrate ($Fe(NO_3)_3·9H_2O$, >98%), and iron (II) chloride tetrahydrate ($FeCl_2·4H_2O$, 98%) were from Alfa Aesar. Iron(III) chloride hexahydrate ($FeCl_3·6H_2O$, >97%) was obtained from PanReac. Fibrinogen (lyophilized powder) was purchased from Merck KGaA (Darmstadt, Germany) with a purity of 98%. Dextran 6 with a molar mass of 6000 g $mol^{-1}$ was obtained from Carl Roth GmbH + Co KG (Karlsruhe, Germany). Polyethylene glycol monomethylether phosphonic acid (PEG-PA; 1900 g $mol^{-1}$) was purchased from Sikémia, France.

Human integrin αIIbβ3 was obtained from Enzyme Research Laboratories (South Bend, USA). 1,2-Dimyristoyl-*sn-glycero*-3-phosphoglycerol, (DMPG; 14:0 PG) and 1,2-dimyristoyl-*sn-glycero*-3-phosphocholine (DMPC; 14:0 PC) were purchased from Avanti Polar Lipids Inc. (Alabaster, USA). SM-2 biobeads were bought from Bio-Rad (Munich, Germany). Tris-Base, ethylenediaminetetraacetic acid (EDTA) and NaCl were supplied by Sigma-Aldrich (Taufkirchen, Germany). $CaCl_2$, $MnCl_2$, Triton X-100 and methanol were purchased from Carl Roth GmbH (Karlsruhe, Germany). Sucrose, phosphate buffered saline (PBS) and sodium dodecyl sulfate (SDS) were obtained from Merck KGaA (Darmstadt, Germany).

### Synthesis and modification of γ-$Fe_2O_3$-nanoparticles

Naked magnetic nanoparticles were prepared according to Massart process.[26,27] In brief, ammonium hydroxide (1 L, 20%) was added to a mixture of ferric and ferrous chlorides (respectively 0.9 mol and 1.8 mol) to obtain magnetite nanoparticles which were oxidized to maghemite by adding an iron nitrate solution (800 mL, 1.3 mol) and heating at 80 °C for 30 min followed by washing and suspension in a nitric acid solution (360 mL, 2 mol $L^{-1}$). After magnetic decantation, 2 L of distilled water and 360 mL of $HNO_3$ at 20%, were added to the solution and the mixture was stirred for 10 min. To obtain the citrated magnetic NPs a fraction of the previous solution was used and sodium citrate at a molar ratio $n$Fe/$n$Cit = 0.13 was added to the nanoparticles and heated up to 80 °C for 30 min to promote absorption of citrate anions onto their surface. This step was followed by magnetic decantation and resuspension in water. The method described by Peng *et al.*[28] was used to coat the surface of the naked NPs with dextran molecules. In brief, 1.5 mL naked maghemite NPs ([Fe] = 17.3 mM) were diluted in sodium hydroxide solution (0.5 mol $L^{-1}$) with an addition of 1 g Dextran 6. This solution was incubated for 5 h in an ultrasonic bath at a temperature below 45 °C. Afterwards, the dextran-modified NPs were purified using magnetic columns by trapping them and washing them several times with deionized water before resuspension in 1.5 mL of ultrapure water. For the surface modification of the magnetic NPs with PEG-PA, 2 mL of the naked $\gamma$-$Fe_2O_3$ NPs were diluted in 15 mL of nitric acid (pH = 2) to avoid their aggregation and mixed with an amount of PEG-PA by considering a density of 1 PEG-PA per $nm^2$.[29] The mixture was ultrasonicated for 30 min followed by stirring overnight. The magnetic NPs were separated on a strong magnet and washed three times with water. Finally, the pH of the suspension was adjusted to 7.0 using 1 M NaOH. The average concentration in nanoparticle samples of the three magnetic suspensions modified by citrate, dextran and PEG, respectively was determined using DLS with multi-angle detection which showed a concentration ~$2 \times 10^{11}$ particles per mL for citrated NPs, ~$5.8 \times 10^{10}$ for dextran-modified NPs and ~$6.5 \times 10^{10}$ particles per mL for PEGylated NPs.

### Fourier transform infrared (FTIR) spectroscopy

The NPs were dropwise air-dried on a glass slide. The dried film was scratched off and loaded on the diamond of the attenuated total reflection (ATR) unit of the Spectrum65 FTIR instrument (PerkinElmer, Waltham, Massachusetts, USA). Spectra were recorded in the range 4000–515 $cm^{-1}$ with 10 scans and a resolution of 4 $cm^{-1}$.

### Protein corona formation

The maghemite nanoparticles with citrate and dextran coatings were diluted 1 : 100 and 1 : 10 in PBS, respectively. 500 µL of fibrinogen (stock concentration of 2 mg $mL^{-1}$) were mixed with 100 µL of the NP solution each (with citrate and dextran) and filled to a total volume of 1 mL with PBS (pH = 7.4). PEGylated NPs were first diluted 1 : 10 in PBS, then a volume of 274 µL was added to the Fb stock (500 µL) and filled to a volume of 1 mL with PBS. All samples were incubated at 37 °C to mimic physiological conditions. NPs with citrate, dextran and PEG coating were analysed before and during incubation with fibrinogen by dynamic light scattering after 0 h, 1 h, 3 h, 5 h, 24 h, 48 h and 72 h.

In addition, after purification with magnetic columns, the bioconjugates were studied using circular dichroism spectroscopy and gel electrophoresis (SDS-PAGE).

### Liposome preparation

To maintain the transmembrane protein under physiological conditions, integrin αIIbβ3 was reconstituted in a lipid membrane following an adapted protocol of Erb and Engel[30] which was used in several studies of integrin αIIbβ3.[18,21,31] In brief, a mixture of 900 nM DMPG : DMPC (1 : 20) was dried under a stream of nitrogen and then under vacuum overnight. Liposome buffer containing 20 mM Tris, 50 mM NaCl and 1 mM $CaCl_2$ was prepared and the pH was adapted to 7.4 with HCl. The lipids were dissolved in liposome buffer containing 0.1% Triton X-100 and 0.2 mg $mL^{-1}$ integrin αIIbβ3 which results in a lipid: protein molar ratio of 1000 : 1. After 2 h incubation at 37 °C, Triton X-100 was removed by adding two times 50 mg SM-2 biobeads for 3.5 h at 37 °C respectively. For separation of non-reconstituted αIIbβ3 from proteoliposomes (*i.e.* liposomes with reconstituted integrin), ultracentrifugation with a four-step sucrose gradient (2 M, 1.2 M, 0.8 M and 0.4 M in liposome buffer) was carried out at 4 °C and 268 000*g* for 24 h. The proteoliposome-containing fraction was harvested and dialyzed against PBS buffer for 72 h using 8 kDa cut-off dialysis cassettes (GE Healthcare, Freiburg, Germany). The vesicles were stored at 4 °C and used for experiments within four days.

### Dynamic light scattering (DLS)

The size determined from the intensity weighted distribution and the zeta potential of the samples were measured with a Zetasizer Nano-ZS/ultra (Malvern Instruments, Kassel, Germany). The represented size data is determined by the intensity weighted distribution. The light scattering caused by nanoparticles is much higher than the scattering of the pure protein and therefore, the pure protein is not detectable in the same measurement. As control, pure integrin in buffer containing 1% Triton X-100 was diluted in PBS to a concentration of 0.4 mg $mL^{-1}$. Triton X-100 was removed by adding SM2-Biobeads followed by dialysis at 4 °C for one day against PBS.

For hydrodynamic diameter measurements, 10 mm-path-length cuvettes (Brand, Wertheim, Germany) were utilized, while for zeta potential determination disposable folded capillary cells (Zetasizer Nano Series, Malvern, Worcestershire, UK) were suitable. Samples were equilibrated 2 min at 37 °C before starting the measurements detecting the intensity of the back-scattered 638 nm-laser beam. The size of four independent samples was determined in automatic mode. To receive information of the electrostatic stability of the colloidal suspensions, the zeta potential of the NP samples was detected within 20 runs and 3 measurements in monomodal mode with a maximum voltage of 10 V.

### SDS-PAGE

Bioconjugate formation was tested by sodium dodecyl sulfate-polyacrylamide gel electrophoresis (SDS-PAGE). For this purpose, samples were incubated in home-built racks equipped with magnets for each tube. The nanoparticles form a pellet during overnight incubation as response to the attached magnet and the supernatant with excessive fibrinogen is removed and the tube refilled with PBS to a volume of 1 mL. This washing step was repeated three times. Afterwards, the pellet was suspended in 20 µL solution of PBS, sample buffer and reducing agent. In the next step, samples were heated at 95 °C for 5 min and then loaded in 8–16% gradient Tris-Glycine gel (Novex, Wedge Well, Invitrogen by Thermo Fisher Scientific, Darmstadt, Germany). All samples were run in duplicate at 225 V for 40 min. Bio-Safe™ Coomassie G-250 stain (Bio-Rad Laboratories Inc.) was used to visualize the protein bands.

### Circular dichroism (CD) spectroscopy

The bioconjugate samples were incubated for 1 h at 37 °C and subsequently purified using magnetic columns (MS MACS, Miltenyi Biotec, Germany). In brief, the magnetic NPs with protein corona were loaded to MS columns (Miltenyi Biotec, Teterow, Germany) which are positioned in a specifically designed strong magnet (MiniMACS Separator, Miltenyi Biotec). While the magnetic nanoparticles were trapped in these columns, most of the free protein was removed with the run-through-fraction. After removing the column from the magnet, the conjugated nanoparticles were resuspended in PBS buffer. A Chirascan CD spectrometer (Applied Photophysics, Leatherhead, UK) was utilized to carry out measurements in the far-UV range. Samples were loaded in 2 mm-pathlength cuvettes (Hellma Analytics, Müllheim, Germany) and spectra were recorded at 25 °C with a scanning time of 1.5 s, a bandwidth of 1 nm and 5 repetitions. All spectra were blank corrected.

### Quartz crystal microbalance with dissipation monitoring (QCM-D)

QCM-D measurements were carried out with a Q-sense Analyzer from Biolin Scientific Holding AB (Västra Frölunda, Sweden) under continuous flow of 25 μL min$^{-1}$ operated by a peristaltic pump (Ismatec IPC-N4, Idex Health & Science GmbH, Wertheim-Mondfeld, Germany) retained at 37 °C. The experiments and cleaning procedures were achieved by a protocol adapted from previous studies and is described by Janke *et al.*[22] QCM-D measures the frequency change ($\Delta f$) and the dissipation change ($\Delta D$) at each measured frequency upon adsorption of material on the sensor surface. The softer the adsorbed layer, the faster the oscillation of the crystal stops, which results in higher dissipation values consequently. Changes in dissipation ($\Delta D$) and frequency ($\Delta f$) of the seventh overtone (35 MHz) are presented in the graphs. After equilibrating the system with PBS buffer for 10 min (ESI Fig. S1 – phase I†), liposomes or proteoliposomes were injected into the system (phase II). After surface adsorption and thus, the formation of a lipid bilayer on the crystal as indicated by a typical peak in dissipation and frequency, the system was washed for at least 10 min with PBS buffer. For integrin activation experiments, PBS buffer containing 1 mM $MnCl_2$, 1 mM $MgCl_2$ and 1 mM $CaCl_2$ was loaded into the system and incubated under continuous flow for approximately 30 min (phase III). NPs or purified bioconjugates (the same purification method as used for CD studies) were added in phase IV followed by rinsing with PBS buffer (phase V). Data analysis was achieved using Q-Tools V.3.0 and QSoft401 V2.5 (both Biolin Scientific AB). $\Delta f_7$ of the NP-injection phase IV was depicted in the graphs and used as an indicator of NP binding efficiency to the lipid bilayer or integrin-containing lipid bilayer.

## Results & discussion

We utilized in this study superparamagnetic γ-$Fe_2O_3$ NPs (atomically represented in pink colour (Fe) and in red (O) in Scheme 1) functionalized with three different ligands citrate, dextran and PEG. These ligands are frequently used in biomedicine because of their properties: surface stabilization at physiological pH, biocompatibility and increased circulation time, respectively.[32–37] A protein corona around the NPs is formed after addition of fibrinogen. The interaction of these functionalized particles bioconjugated with the blood protein fibrinogen and lipid membranes containing integrin (non-activated and activated) were investigated using QCM-D. Furthermore, different biophysical tools (FTIR, DLS, SDS-PAGE, CD spectroscopy) were applied to characterize the NPs, the lipid membranes and the corresponding proteins.

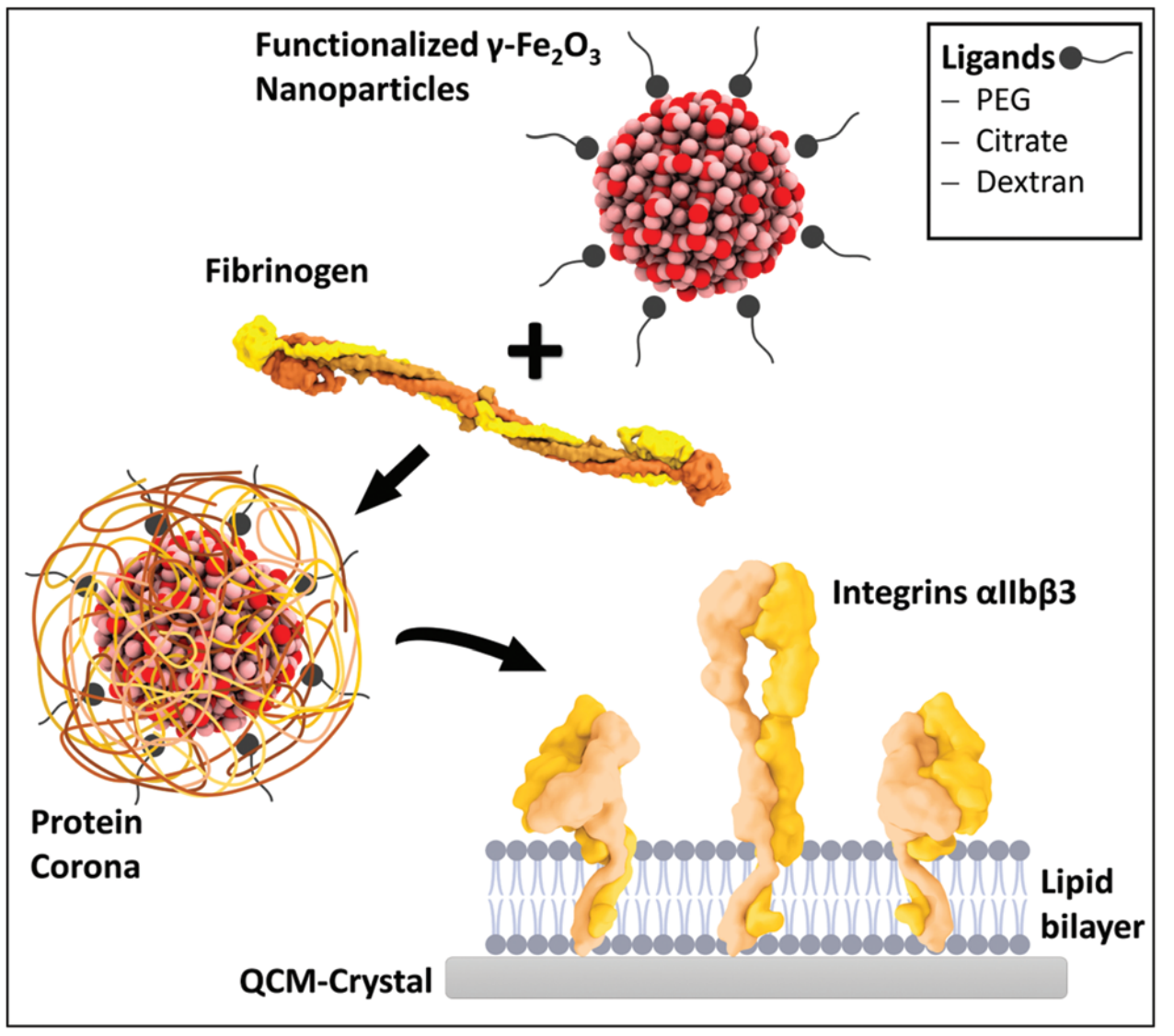

Scheme 1 Overview of this study, showing the NP functionalization with different ligands. The NPs formed a protein corona with fibrinogen and the interaction of these particles with lipid membranes containing integrin were studied using QCM-D.

### Nanoparticle characterization

Modified nanoparticles were analysed by FTIR spectroscopy to verify the grafting of the different ligands on the surface of NPs. Here, the spectra of the naked γ-$Fe_2O_3$ NPs, the different coating molecules as well as the functionalized γ-$Fe_2O_3$ NPs are displayed in Fig. 1. The FTIR spectrum of the naked magnetic nanoparticles as well all the modified ones showed the characteristic absorption bands in the 630–550 cm$^{-1}$ range attributed to the vibrations of Fe–O bands.[38] After citrate modifications, the citrated nanoparticles (γ-$Fe_2O_3$-Cit) displayed the representative vibrations of trisodium citrate at 1570 cm$^{-1}$ and 1385 cm$^{-1}$ assigned to C═O and C–H bending vibrations.[39] In addition, at the positions 1253 cm$^{-1}$ and 1073 cm$^{-1}$ the bands for C–O asymmetrical stretching of citrate were observed.[39] After surface modification of the naked NPs with dextran new absorption bands appeared at 3387 cm$^{-1}$ relative to the hydroxyl groups of the molecule, and at 1150 and 1010 cm$^{-1}$ corresponding to C–O deformations as well at 916, 852 and 764 cm$^{-1}$ relative to the α-glucopyranose of dextran.[40–42] FTIR spectrum of PEG-PA modified NPs ($\gamma$-$Fe_2O_3$-PEG) presented the characteristic PEG vibrational bands located at 2878 $cm^{-1}$ (C–H stretching), 1098 $cm^{-1}$ (C–O stretching) as well as positioned at 960 $cm^{-1}$ and 842 $cm^{-1}$ (C–H rocking) which slightly shifted compared to free PEGylated molecules.[43,44] In addition stretching vibrations near 1060 $cm^{-1}$ which could be assigned to P–O–C and at 1241 $cm^{-1}$ referring to P═O were observed showing clearly the successful grafting of PEG-PA on the surface of magnetic NPs.[45]

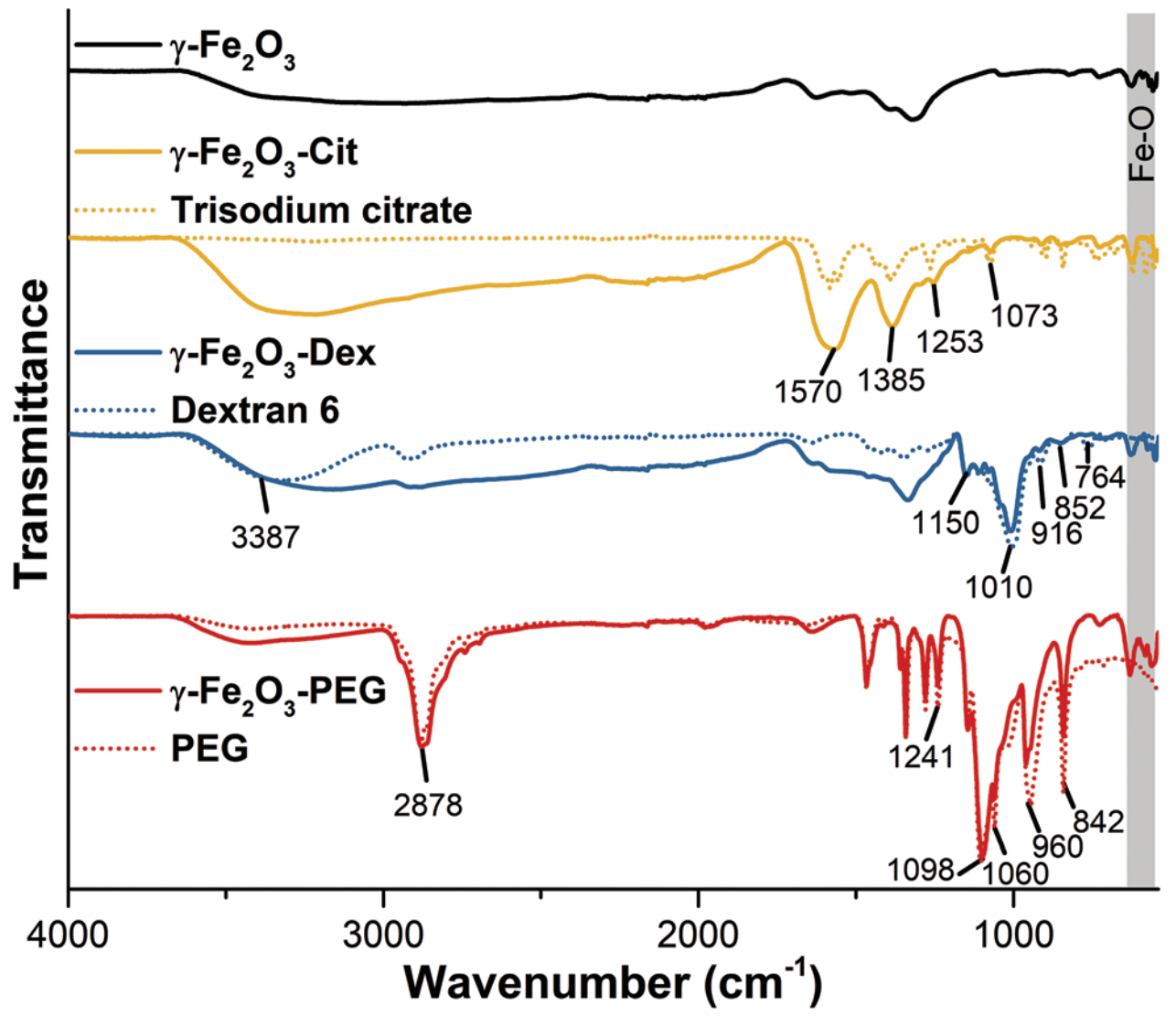

**Fig. 1** FTIR spectra of dried naked $\gamma$-$Fe_2O_3$-NPs suspensions (black) and those functionalized with citrate (yellow), dextran (blue) and PEG (red) as well as the corresponding coating material (dotted lines), respectively. Data is monitored in ATR-mode and baseline-corrected. Here, the spectra are separated for a better overview. The characteristic peak positions are indicated by the corresponding wavenumber values.

A TEM image of the magnetic NPs used in this study is presented in Fig. 2A. It shows that the nanoparticles have rock-like shapes with an average diameter of 10.2 ± 0.11 nm. The hydrodynamic diameter of the nanoparticles calculated from the intensity weighted distribution of the colloidal suspensions are shown in Fig. 2B. Before mixing with Fb, citrate-stabilized particles revealed a size of ~35 nm, while the dextran-modified and the PEG modified particles showed a size of ~38 nm and 42 nm respectively. However, the size of all non-conjugated nanoparticles increased over time, but more dramatically for the dextran modified NPs reaching ~120 nm which resulted after 5 h in the formation of aggregates and their precipitation followed by the citrate and PEG-modified NPs. These results show the superiority of PEG as a coating and stabilising agent in PBS buffer.

### Fibrinogen corona formation

Fibrinogen was used for bioconjugation experiments because it is one of the most abundant blood plasma proteins whose binding was verified to numerous types of NPs.[46] Even in experiments with competitive proteins also abundant in the blood plasma *i.e.* human serum albumin and transferrin, the adsorption tendency for fibrinogen was found to be higher.[17] After exposing NPs to plasma, the protein corona generally consists of fibrinogen.[47,48]

During incubation of Fb with each of the surface modified magnetic NPs the resulting hydrodynamic diameter was measured with time (grey shaded bars in Fig. 2B). After a few minutes a significant size increase for citrated NPs and for dextran modified NPs to ~75 nm and 100 nm respectively was observed demonstrating clearly the fast interaction of Fb with the NPs and the formation of bioconjugates. Citrated and dextran modified NPs reached stable size values of 66 nm and 70 nm respectively after one hour, which remained constant even after 72 h. In case of the PEGylated particles, the behaviour was different from citrate and dextran coated magnetic NPs. In presence of Fb the size did not change significantly (~42 nm) within the first minutes. However, the size increased continuously within the studied time range up to 90 nm after 72 h in contrary to the fast size stabilization obtained after one hour of incubation for the citrated and dextran coated ones. These results show clearly the role of PEGylation in increasing the colloidal stability of the NPs by delaying their surface from opsonization due to steric hindrance.[36,49] The interaction of Fb with the NPs was also studied by measuring the zeta potential before and after incubation of the NPs with Fb for 3 h (Fig. 2C). The formation of bioconjugates resulted in a decrease of the zeta potential from −38 mV to −2 mV for the citrated NPs and from −37 mV to −7 mV for the dextran modified NPs whereas for the PEGylated NPs the zeta potential decreased slightly from −6 mV to −4 mV. Taking into account that in PBS buffer the protein Fb is expected to be negatively charged because of its isoelectric point of ~5.5,[50] the bioconjugate formation with all NPs is expected to increase the negative zeta potential. We measured the zeta potential of Fb in PBS which was near zero. Consequently, taken together these results suggest a strong binding of Fb to the surface of the magnetic NPs. The adsorption of Fb onto the surface displaces the citrate and dextran, which are weak binding ligands compared to phosphonic acid that binds strongly through bidentate or tridentate Fe–O–P coordination bonds.[51] Fb is then adsorbed on their surfaces and the zeta potential values highly decrease. Moreover, considering the zeta potential and the size measurements from DLS, the results indicate that Fb protein can act as a surface coating molecule providing colloidal stability of the magnetic NPs instead of the original coatings. In the case of the PEG-modified NPs the adsorption of Fb is more delayed probably due to the high density of grafting of the PEG moieties and to the strong binding and complexation of the surface of the magnetic NPs by the phosphonic acid group. These results agree with those obtained from DLS measurements, showing for citrate- and dextran-NPs a faster formation of larger bioconjugates than for PEGylated NPs.

Furthermore, the formed protein corona around the different modified $\gamma$-$Fe_2O_3$-NPs was estimated by SDS-PAGE. The bioconjugates were purified by applying a static magnetic field overnight, which resulted in pellet formation and thus, enabled the washing process. Fig. 2D depicts the Coomassie-stained gel revealing the protein adsorption to the different modified NPs. The first lane shows the protein marker. Fibrinogen displays three bands on the gel, attributed to Aα- (~67 kDa), Bβ- (~55 kDa) and γ-chain (~48 kDa).[52] As expected, citrated-, dextran-coated and PEGylated NPs showed no protein bands, while strong bands were observable for the bioconjugates. Obviously, the Fb bands for the bioconjugates with PEGylated NPs were less intense revealing a much lower protein concentration than for citrated and dextran-modified particles in agreement with the results obtained by DLS and zeta measurements. In view of these results and taking into account the high density grafting of PEG (1 PEG-PA per $nm^2$), we can hypothesize that Fb is intercalating in the extended PEG brushes of ~5.7 nm length.[29] Furthermore, the protein bands of the bioconjugates showed a much higher intensity than the positive controls with 200 μg $mL^{-1}$ and 100 μg $mL^{-1}$ Fb. While the initial Fb concentration was 1 mg $mL^{-1}$ the purified samples still consist of high amounts (>200 μg $mL^{-1}$) of the protein after three washing steps which indicates the enrichment of NPs with Fb and thus, a strong affinity. Furthermore, we determined comparable protein concentrations of samples purified with magnetic columns using a Bicinchoninic Acid Kit (Sigma-Aldrich, Steinheim, Germany) that revealed for citrated NPs Fb concentrations of 205 μg $mL^{-1}$ (±21 μg $mL^{-1}$), for dextran-coated NPs 290 μg $mL^{-1}$ (±21 μg $mL^{-1}$) and for PEGylated NPs the protein concentration was below the detection limit of 200 μg $mL^{-1}$.

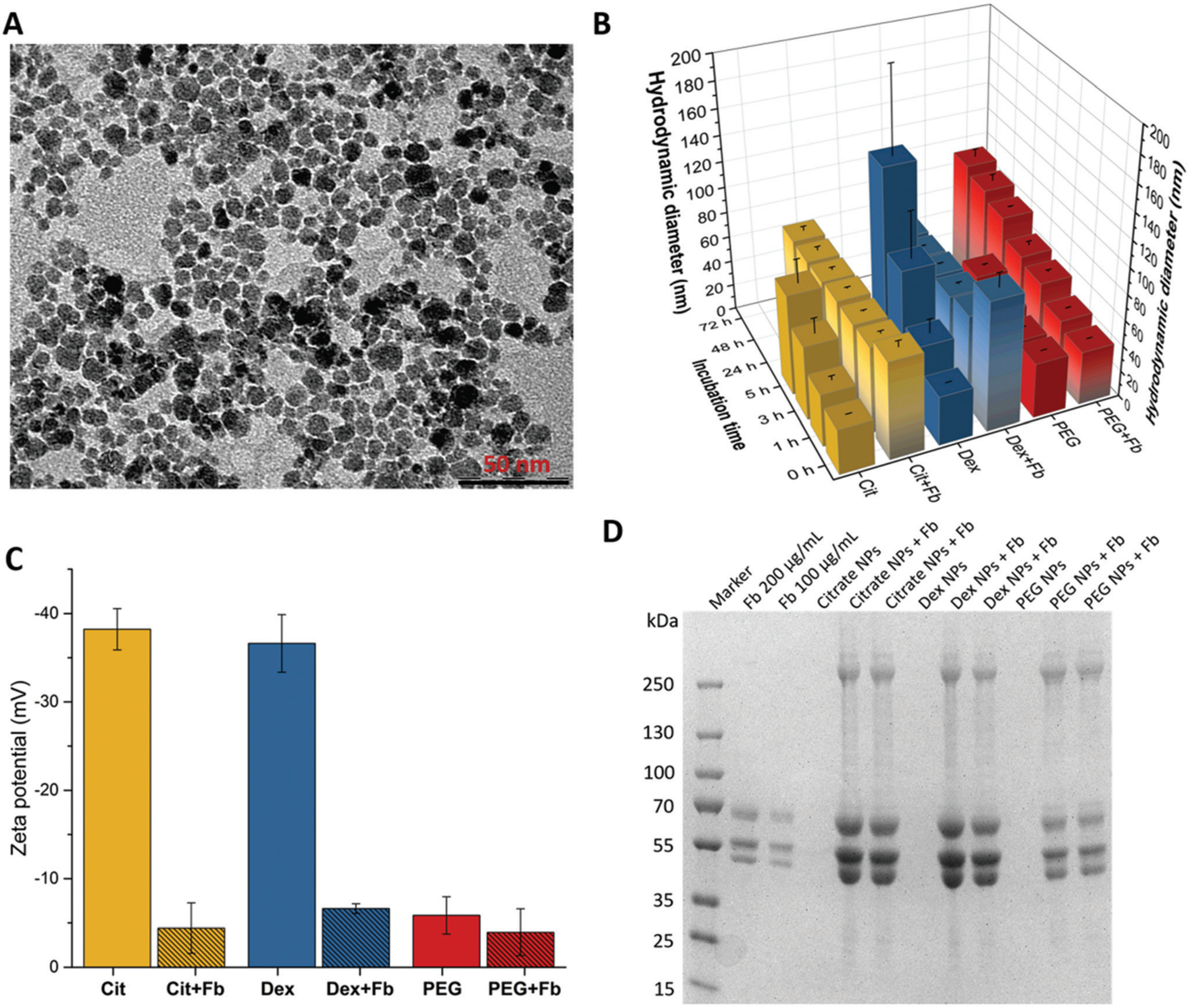

Fig. 2 (A) TEM image of the magnetic nanoparticles showing their rock-like shape, (B) hydrodynamic diameter (intensity weighted distribution) after 0 h, 1 h, 3 h, 5 h, 24 h, 48 h and 72 h of incubation with Fb in PBS. The missing bars show the conditions where the particles were aggregated. (C) Zeta potential of citrate- (yellow), dextran- (blue) and PEG-functionalized (red) maghemite nanoparticles and their bioconjugates with fibrinogen in PBS after 3 h. (D) Coomassie-stained reducing Tris-Glycine gel (8–16%) showing the protein content of the different bioconjugates with citrated, dextran-coated and PEGylated nanoparticles by magnetic purification.

In addition, the SDS gel shows a fourth band with a high molecular weight for the bioconjugate samples, which could be attributed to the non-reduced protein. All these data confirm the effective bioconjugation with fibrinogen for citrated, dextran- and PEG-coated maghemite NPs.

Moreover, we examined the secondary structure of Fb with the aim to identify if any possible structural changes would occur due to bioconjugation. Those variations may lead to various membrane interaction *e.g.* a protein–protein interaction with integrin. The CD data measured for the different modified maghemite NPs with citrate, dextran and PEG and their bioconjugates with Fb after 1 h incubation and subsequent magnetic purification is illustrated in ESI Fig. S2.† In summary, changes in the spectra of fibrinogen are attributed to the background signal of the nanoparticles and are not caused by structural changes of the protein due to bioconjugation. These results show that the coatings and the interaction of the protein with the surface of the magnetic NPs have no influence on the structure of the protein.

Previous literature reports on conformational changes of Fb through the interaction of different NPs *e.g.* zeolith and gold nanoparticles.[53–57] Derakhshankhah *et al.*[55] showed that Fb is bound to zeolith NPs due to strong electrostatic interaction and enhanced structural changes of fibrinogen with increasing NP concentration. Molecular dynamic simulations with gold NPs also verified structural changes of Fb due to NP binding which are in favour with inflammation response. They pointed out that the application of gold NPs might be harmful and even the coatings may be degraded.[56] In addition, it was also shown that $\gamma$-$Fe_2O_3$ NPs induce conformational changes of Fb.[46] In comparison to the study by Zhang *et al.*, our investigated bioconjugates are purified using magnetic separation and do not show structural changes as presented by our CD data.

We aimed to elaborate a biomimetic model which can allow to study the interaction of the coated NPs as well as their bioconjugates with membranes and the receptor αIIbβ3. Liposomes were formed as described in the Experimental section followed by the integration of the protein αIIbβ3 in their membrane. To validate the successful integrin αIIbβ3 reconstitution into liposomes, DLS measurements were first carried out (Fig. 3A).

The hydrodynamic diameter of the proteoliposomes (531 nm) was larger than that of unmodified liposomes (423 nm). Furthermore, an additional smaller peak appeared at 90 nm for both types of liposomes, due to sample polydispersity resulting from the liposome preparation procedure. Control of pure integrin is also shown. Removal of Triton X-100 leads to aggregation of pure protein sample resulting in a hydrodynamic diameter of 35.5 nm, whereas for pure integrin diluted in buffer containing detergent a size of around 20 nm is expected. The presence of the αIIb- and β3-subunits reconstituted in the liposomes was also verified by a denaturizing SDS-PAGE (Fig. 3B). Both subunits migrated as two visible bands at ~115 kDa (blue arrow) and ~90 kDa (red arrow) in the proteoliposome sample, but not in the liposome sample.

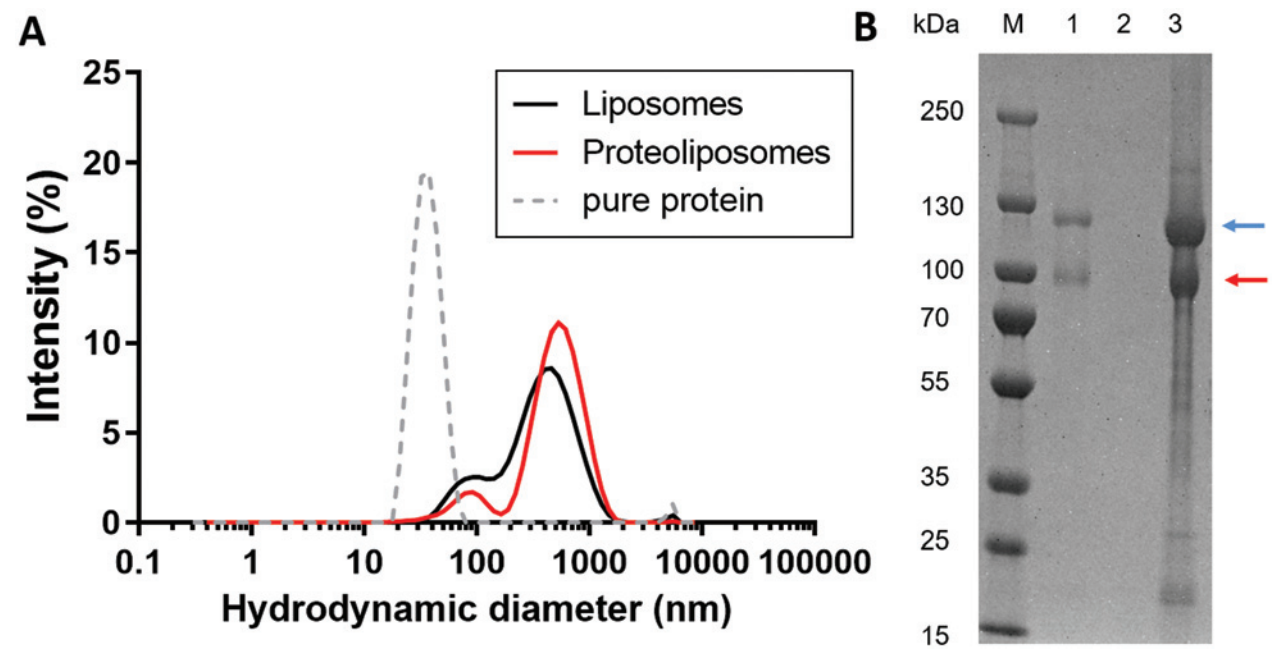

Fig. 3 (A) DLS data showing the hydrodynamic diameter of liposomes (black) and proteoliposomes (red) averaged over three data sets in PBS at 37 °C. Additionally, the hydrodynamic diameter of pure integrin (grey dashed line) is shown. (B) Reductive SDS-PAGE of proteoliposomes (1), liposomes (2), pure integrin protein with a concentration of 0.2 mg $mL^{-1}$ (3) and protein molecular weight standard (M). The bands corresponding to the αIIb- (blue) and β3-subunit (red) are indicated by arrows.

Thus, the successful reconstitution of integrin αIIbβ3 into liposomes was demonstrated by DLS measurements and SDS-PAGE.[30] In addition, the reconstitution was verified in our previous studies[22] using transmission electron microscopy (TEM) as imaging techniques which showed spherical proteoliposomes with globular heads and stalk domains of the ectodomain of integrin.

### Monitoring the interaction of Fb–NP bioconjugates with artificial membranes by QCM-D

The interaction of the different coated magnetic nanoparticles with artificial cell membrane systems (lipid bilayers) was studied by QCM-D which allows the detection of mass changes and differences in viscoelastic properties in real-time.[58]

In particular, we analysed the behaviour of citrate-, dextran- and PEG-coated maghemite nanoparticles before and after conjugation with fibrinogen upon interaction with an artificial lipid membrane containing integrin. On a silicon dioxide sensor surface, liposomes tend to form a supported lipid bilayer (SLB) *via* vesicle fusion technique. This technique makes the analysis of membrane proteins and their interaction behaviour in a biomimetic system feasible.[58–60]

Compared to blank liposomes, the injection of the proteoliposomes (ESI Fig. S1† – phase II†) lead to a strong increase in the dissipation ($D$), with a frequency ($f$) which was only slightly higher. However, the increase in $D$ is not as high as expected for a vesicle layer, which gives evidence for a SLB with higher dissipation due to remaining liposomes on the SLB[61] and the huge ectodomain of the integrin.[62]

After formation of the SLB, unconjugated or Fb-conjugated NPs with different coatings were run over the sensor chip. Fig. 4A displays the equilibrium values of the frequency and dissipation shift measured between the time of injection of the NP and the time of rinsing with the buffer (ESI Fig. S1 – start of phase IV and V†). These values correspond to the adsorbed mass of NPs during that time period. Exposure of the SLB to a dispersion of non-conjugated NPs lead to no sig-nificant changes in f and D for all types of NPs (Fig. 4). There is a minor tendency of dextran- and PEG-coated nanoparticles to adsorb to the SLB ($\Delta f \sim 3$ Hz and $\Delta D \sim 1 \times 10^6$). Hence, no significant interactions between the SLB and unconjugated- as well as Fb-conjugated NPs were observed. Interestingly, for the SLB with integrins, the results show substantial changes up to 33 Hz in $f$ and $6 \times 10^{-6}$ in $D$ upon injection of dextran- and citrate-coated NPs. However, for PEG-coated NPs only tenfold less binding (3 Hz in $f$ and $0.5 \times 10^{-6}$ in $D$) was reached.

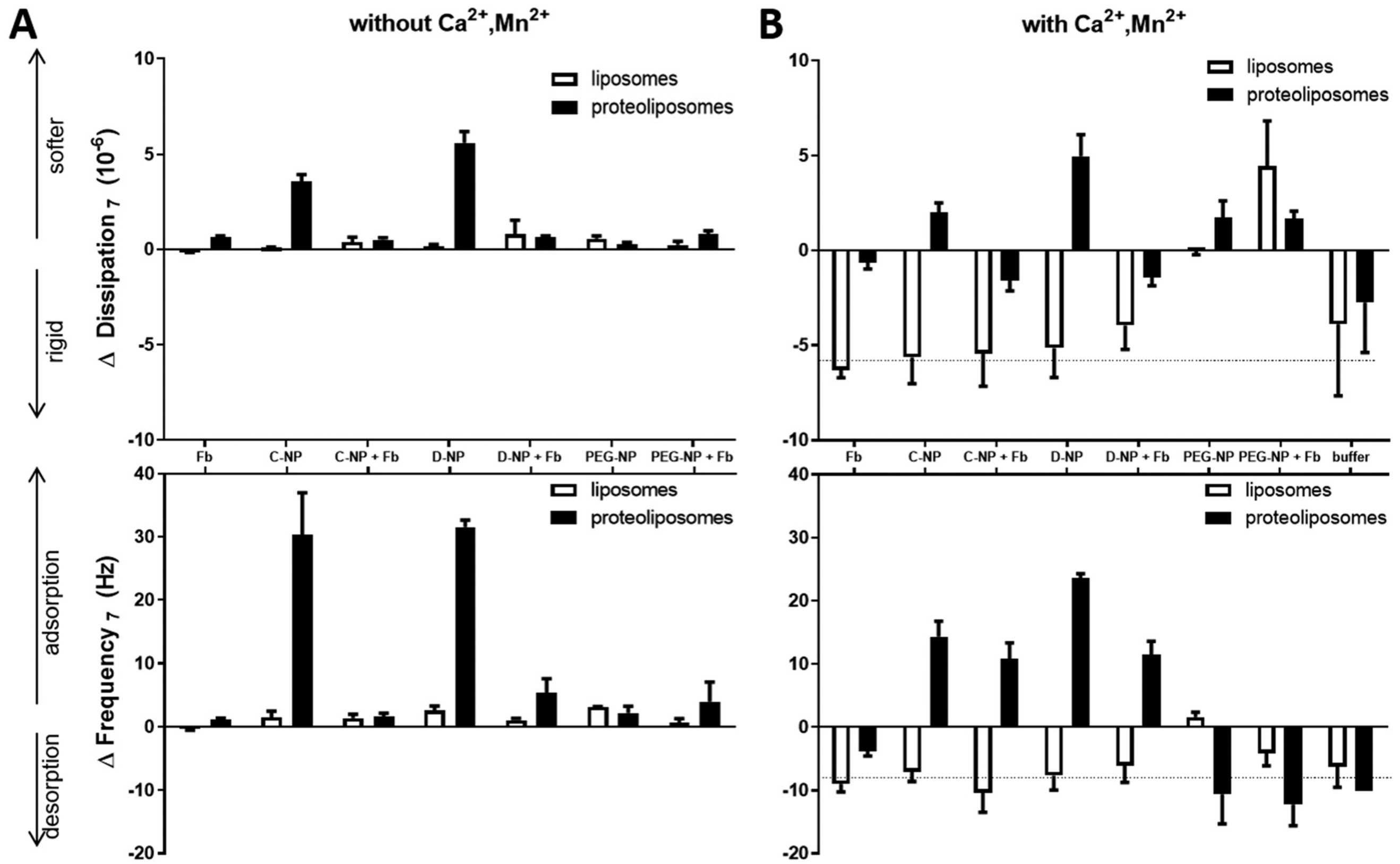

Fig. 4 (A) Changes in dissipation (top) and frequency (bottom) during injection of fibrinogen (Fb), naked NP (C-citrate, D-dextran, PEG-NP) and fibrinogen-coated NPs (+Fb) obtained from QCM-D experiments during experimental phase IV (ESI Fig. S1†). NPs and bioconjugates were pumped over SLB without (white) or with incorporated integrin (black). The running buffer was PBS without any divalent ions. (B) Changes in dissipation (top) and frequency (bottom) during injection of fibrinogen (Fb), naked (C-citrate, D-dextran, PEG-NP) and fibrinogen-coated NPs (+Fb) after treatment with divalent ions, obtained from QCM-D experiments during experimental phase IV. NPs and bioconjugates were run over SLB without (white) or with (black) incorporated integrin. The running buffer was PBS with 1 mM $CaCl_2$, $MgCl_2$, $MnCl_2$.

PEGylated NPs conjugated to Fb as well all the coated magnetic NPs without conjugation showed comparable changes in $f$ and $D$ in response to the integrin-containing SLB. These results confirm in accordance with the results from DLS, that the amount of attached Fb to the PEG coated magnetic NPs is much lower than for the citrated and dextran modified NPs. The bound Fb concentration might be low to produce reliable changes in QCM-D signals. Again, the lack of interactions between PEG and integrin supports the fact that PEGylation acts as an effective steric hindrance for binding to blood components which also protects from clearance by the immune system and thus, increases the circulation time.[37]

However, citrate- and dextran-coated NPs with Fb corona showed a significantly lower binding tendency (~5 Hz in $f$ and $0.5 \times 10^{-6}$ in $D$) to the lipid bilayer compared to their relative NPs without Fb. Nevertheless, citrate- and dextran-coated NPs with Fb as well as Fb alone show a similar binding tendency for both, SLB and integrin-containing SLB.

The formed Fb-corona results in reduced interactions between the NP with the SLBs. The coverage of the surface and thus, the reduction of surface charge leads to the reduction of electrostatic interactions as well as unspecific binding. Similar results were already observed for other types of NPs.[10,59] However, non-specific and specific interactions involve a large set of adhesion forces *e.g.* electrostatic, hydrophobic, van der Waals forces and ligand–receptor binding.[14,63] NPs may bind to the integrin inserted in the lipid bilayer by electrostatic effects upon absorption.[64] Generally, Fb corona formation of all NPs showed an increased tendency of binding to the integrin-containing lipid bilayer compared to the lipid bilayer without integrin due to potential binding sites on the Fb for the integrin αIIbβ3.[19]

Here, it has to be mentioned that also the number of injected particles potentially affects the binding tendency and kinetic to the membrane. The Fb concentration in human plasma is about 2.0–4.5 mg $mL^{-1}$. The Fb concentration in our control experiments (only Fb, without NPs) was 1 mg $mL^{-1}$ which corresponds to the initial concentration of bioconjugate preparation, while the protein concentration can be assumed to be less after magnetic purification of the Fb–NP-complexes, which is much lower than the physiological concentration However, 1 mg $mL^{-1}$ Fb in our control experiments shows comparable binding behaviour than the less concentrated Fb-coated NPs.

### Interaction of Fb–NP bioconjugates with activated integrin in lipid membranes

Since integrin dynamics and stability depend on several divalent ions,[20,21] the lipid bilayers were treated with buffer containing $Ca^{2+}$, $Mg^{2+}$ and $Mn^{2+}$, where the latter is known to favour the active state of αIIbβ3 externally. With these conditions, the activation of platelets can be simulated in a biomimetic system with biophysical tools upon contact with maghemite NPs conjugated with Fb. After treatment with divalent ions ($Ca^{2+}$, $Mg^{2+}$, $Mn^{2+}$) of both SLB and integrin-containing SLB (Fig. 4B), $\Delta f$ and $\Delta D$ upon NPs injection behave differently compared to non-treated layers (Fig. 4A). The first striking variance is that even the buffer switch from divalent ion-containing PBS to normal PBS (Fig. 5) showed changes in $f$ (~ −10 Hz) and $D$ (~ $-6 \times 10^{-6}$) for both, liposome and proteoliposome derived SLB, but not for control buffer injections on blank $SiO_2$ surface. Those negative changes in $f$ and $D$ indicate besides mass desorption, a more rigid and thinner surface layer compared to the divalent ions treated layer. Hence, the mentioned ions lead, by binding to the lipid headgroups, to a swelling of the water layer between the bilayers, which was already observed in other studies.[65,66] Please note, due to stability issues, all NPs and their bioconjugates were suspended in PBS without divalent ions. In fact, after the treatment of the SLB with buffer containing divalent ions, NPs or bioconjugate suspensions without divalent ions were injected. Consequently, the buffer switch leads to a mismatch resulting in a shift in frequency and dissipation upon NP injection. Therefore, results in that signal range are hardly evaluable and have to be interpreted with care. However, divalent ions have a significant effect on SLB as well as on integrin-containing SLB.

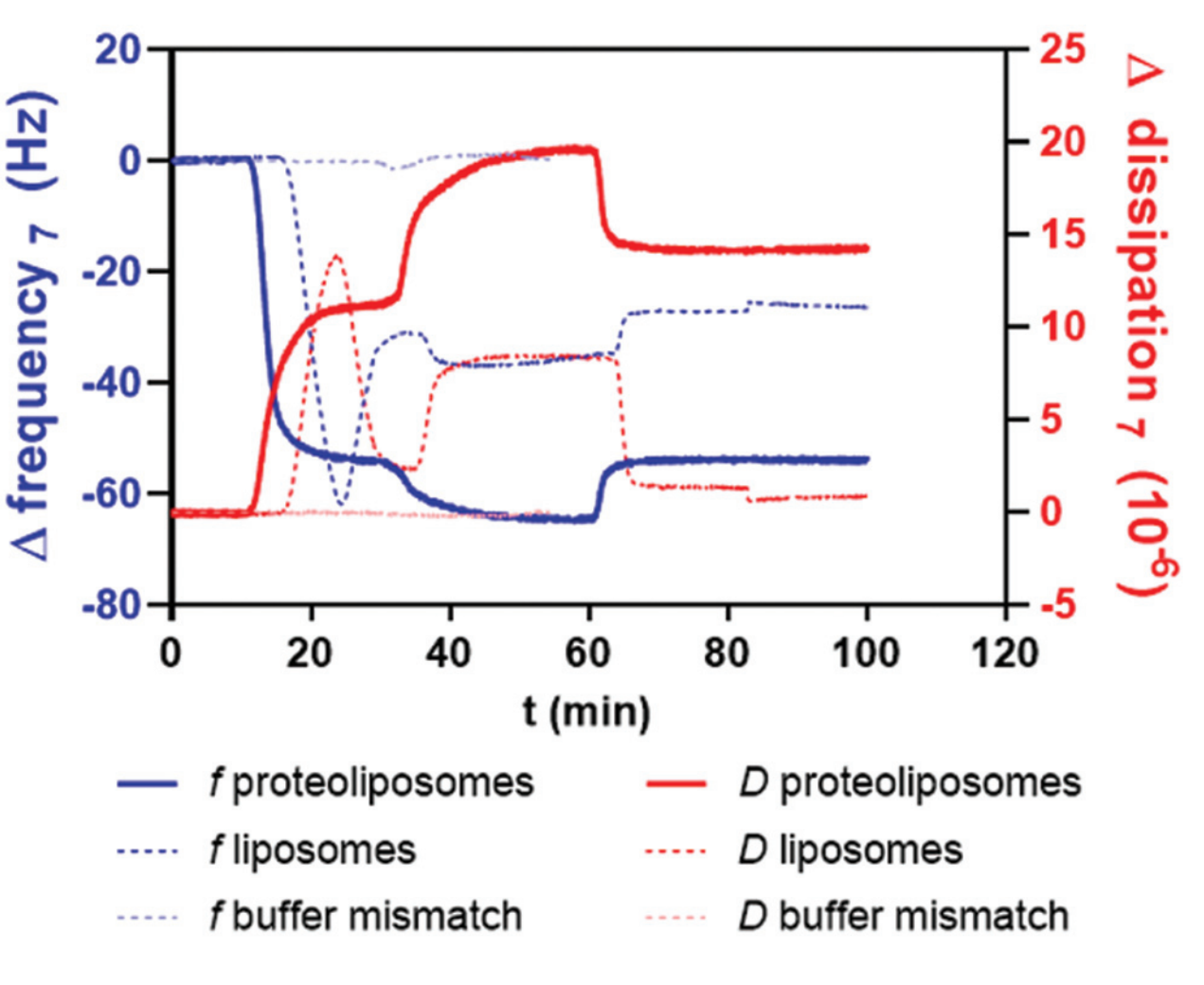

**Fig. 5** Representative QCM-D experimental profile for the buffer changes. Changes in dissipation (D-red) and frequency $f$ (blue) of the seventh overtone at 37 °C. PBS buffer was injected over the $SiO_2$ sensors and after reaching a baseline, liposomes or proteoliposomes were injected (10 min) and the formation of a bilayer was observed except for control injections of buffer on blank surface. After a washing step with PBS containing 1 mM $Mn^{2+}$, 1 mM $Ca^{2+}$, 1 mM $Mg^{2+}$ (30–60 min). Rinsing with PBS buffer followed for bilayer samples. Buffer mismatch on blank surface is indicated in light blue ($\Delta f$) and light red ($\Delta D$).

The liposome-derived lipid bilayer treated with divalent ions revealed comparable results to the buffer control during NP injection. However, the injection of citrated and dextran NPs as well as the injection of their bioconjugated counterparts to the integrin-SLB resulted in a signal that exceeded the signal shift in f during buffer mismatch. As demonstrated in experiments without any divalent ions, treatment with surface-coated NPs equally gives the highest $\Delta f$ (25 Hz) and $\Delta D$ ($6 \times 10^{-6}$) for dextran-coated NPs upon binding to divalent ion-treated integrin-SLB and thus, overcome the buffer mismatch effect. Fb-conjugated NPs with citrate and dextran modifications reach $\Delta f$ of 12 Hz while $D$ decreases to $2 \times 10^{-6}$ simultaneously which is compared to the non-treated integrin a multiplied response in signal. These data verify the assumption of stabilization and activation of αIIbβ3 through treatment with divalent ions. Consequently, the bioconjugates of NPs and Fb show an increased binding affinity. The integrin head domain opens up upon $Mn^{2+}$ treatment, which was shown in other publications,[19,22] and is able to bind more of fibrinogen-coated NPs *via* the RGD motif in the Aα-chain or by the KQAGDV sequence located in the γ-chain compared to non-treated bilayers.

In contrast, PEG-modified NPs as well as Fb-coated PEG NPs show no changes in $f$ beyond the buffer mismatch, but an increase in $D$ of $2 \times 10$ for integrin-containing bilayer and even $5 \times 10^6$ for the interaction with blank lipid bilayer. Only minor adsorption of PEGylated NPs to the lipid bilayer or integrin-containing bilayer after treatment with divalent ions is detected, while the viscoelasticity was increased upon PEG-NPs injection, which can be due to hygroscopic effect of the PEG-polymer increasing the softness of the adsorbed layer.[67]

Summarized, Fb bound to dextran- and citrate-coated NPs interact with integrin-containing lipid bilayer especially upon treatment with divalent ions, whereas PEG-coating reveals minor interaction. These results confirm the work of Suk *et al.*,[37] where PEG-coating inhibit NP binding to components of the blood stream. Some studies showed that Fb-coated NPs are involved in host response, activation of immune cells and platelet aggregation.[53,68] However, there are contradictory results concerning the influence of maghemite NPs on the coagulation. Iron oxide NPs treatment of rats prolongs the coagulation time of platelets,[69] while other groups show enhanced platelet aggregation upon treatment with maghemite NPs.[70,71]

## Conclusions

In this paper we firstly investigated the interaction of citrate-, dextran- and PEG-modified magnetic nanoparticles with fibrinogen, an abundant plasma protein. Based on DLS, zeta potential and SDS-PAGE, we showed that citrate and dextran are easily displaced by Fb, which upon binding to the surface stabilize the magnetic NPs from aggregation over time. The PEG coating increased the stability of the magnetic NPs by acting as a barrier for Fb adsorption on the surface of the NPs. Circular dichroism of Fb–NPs bioconjugates showed no effect of the ligands or the magnetic NPs on Fb structure. Moreover, we introduced a biophysical platform to investigate NP interaction with αIIbβ3 receptor-containing lipid bilayers that serves as an artificial platelet membrane model. We showed that the coating of the NPs as well as their resulting bioconjugates with Fb significantly influence the interaction with membranes containing receptors such as αIIbβ3. While PEG-coating seems to shield this interaction, which possibly eliminates the toxicity and immunogenicity of NPs in the body, the Fb conjugates present a much higher binding to the integrin-containing membrane. The coating of NPs with fibrinogen together with other substances *e.g.* drugs might be applicable to enhance the targeted interaction with membrane proteins. Thus, this study provides new insights in focus of biomedical applications. Additional experiments are needed to shed more light on the intermediate steps between NP characteristics and biological assays. Particularly, studies which involve proteoliposomes are essential to explore the effect of surface curvature on NP-membrane interactions.

## Conflicts of interest

There are no conflicts to declare.

## Acknowledgements

We acknowledge the support of Norman Geist (Biophysical Chemistry group of Prof. M. Delcea, Greifswald, Germany) who prepared Scheme 1 of this manuscript. The financial support by the Federal Ministry of Education and Research (BMBF) within the NanoImmun Project (FKZ 03Z22C51) and the European Research Council (ERC) Starting Grant "PredicTOOL" (637877) to MD is gratefully acknowledged.